\documentstyle[12pt]{article}

\font\tenrm=cmr10
\font\tenit=cmti10
 1
\font\elevenrm=cmr10 scaled\magstep 1
 1

\def\beq{\begin{equation}}
\def\eeq{\end{equation}}

\renewenvironment{thebibliography}[1]
 { \elevenrm
   \begin{list}{\arabic{enumi}.}
    {\usecounter{enumi} \setlength{\parsep}{0pt}
     \setlength{\itemsep}{3pt} \settowidth{\labelwidth}{#1.}
     \sloppy
    }}{\end{list}}

\begin{document}

\begin{titlepage}

\rightline{\vbox{\halign{&#\hfil\cr
&NTUTH-96-03\cr
&February 1996\cr}}}
\vspace{0.2in}

\begin{center}
\vfill
        {\bf POSSIBLE SOLUTION TO \boldmath{$R_b$ -- $R_c$} PROBLEM                \footnote{Talk presented at II Rencontres du Vietnam, Ho Chih Minh City,
October 21 -- 28, 1996.
Work done in collaboration with Gautam Bhattacharyya
                                                                                    and Gustavo Branco.}\\}

\vfill
{\tenrm GEORGE WEI-SHU HOU\\}
\baselineskip=13pt
{\tenit Department of Physics, National Taiwan University\\}
\baselineskip=12pt
{\tenit Taipei, Taiwan 10764, R.O.C.\\}
\end{center}
\vfill

\begin{abstract}
We investigate the $R_b$--$R_c$ problem,
starting from the more diffcult $R_c$.
Introducing an isosinglet charge 2/3 quark $Q$ moves both
$R_c$ and $R_b$ in the right direction.
If one allows for large $c$-$Q$ {\it  and} $t$-$Q$ mixings,
then $R_c$ could be reduced because of singlet content of charm quark,
while $R_b$ gets enhanced by a light effective top quark mass in the loop.
The heavy quark observed at CDF would be dominantly a singlet quark,
while the top quark is lighter than $M_W$.
It is necessary to introduce a second Higgs doublet,
where $H^+$ is heavy while 
at least one exotic neutral Higgs boson is very light.
$H^+$--$h^0$ splitting accounts for $\delta\rho$,
while light $h^0$
induce fast $t\to c + h^0$ decay and 
hides the actual top quark at the Tevatron.
The scenario can be immediately tested at
LEP II via search for light top production (and toponium!).
At Tevatron, one should search for exotic decay modes
such as $Q\to Z + X$ or $Q\to H + X$,
or measure the BR for the standard $bW$ mode.
Light top search should also be renewed.
\end{abstract}
\vfill
\end{titlepage}

{\bf\noindent 1. Introduction}
\vglue 0.1cm
\baselineskip=14pt

We have heard from Martin$^{1{\scriptsize)}}$ 
and Altarelli$^{2{\scriptsize )}}$ on the
experimental and theoretical aspects of the $R_b$ -- $R_c$ problem.
Rather than showing the offending ``99.9\% C.L." figure,
let us give the result from the
multiple parameter fit,
\begin{equation}
R_b^{\rm exp}  =  0.2219 \pm 0.0017, \ \ \ \
R_c^{\rm exp}  =  0.1543 \pm 0.0074,
\end{equation}
vs.  Standard Model (SM) values (for $m_t = 180$ GeV)
$R_b^{\rm SM}  =  0.2156$, and
$R_c^{\rm SM}  =  0.172$.
Thus,
\begin{equation}
\delta R_b  \simeq  + 3.7 \sigma, \ \ \ \
\delta R_c  \simeq  - 2.5 \sigma.
\end{equation}
From both experimental and theoretical considerations,
{\it nobody} seems to like it this way (especially $R_c$).
But, {\it what if}?
We shall give here an {\it ad hoc} ({\it \`a la} Altarelli,) but perhaps natural
solution$^{3{\scriptsize)}}$ that offers 
absolutely tantalizing phenomenology in the near term.

\vglue 0.4cm
{\bf\noindent 2. Conservative approach for \boldmath{$R_c$}}
\vglue 0.1cm

Consider the minimal extension of adding just one
charge $+2/3$ isosinglet quark $Q$.
One then has new gauge invariant mass terms
$\bar{Q}_L Q_R$ and $\bar{Q}_L u_{jR}$,
and Yukawa coupling terms $\bar{u}_{iL} Q_R$,
where $u_i$ denotes standard up-type quarks.
As a result, $Q$ mixes with $u$, $c$ and $t$.
For sake of discussion, let us ignore 1st generation and set 
$V_{\rm KM} \equiv I$ (i.e. discuss only ``Cabibbo allowed" modes).
One then has the charged current
\begin{equation}
(\bar c_L\ \bar t_L\ \bar Q_L)
          \left( \begin{array}{cc}
                    C_2 & 0 \\ -S_2S_3 & C_3 \\ +S_2C_3 & S_3
          \end{array} \right) \gamma_\mu
          \left( \begin{array}{r}
                    s_L \\ b_L
          \end{array} \right),
\end{equation}
where $S_i \equiv \sin\theta_i$, $C_i \equiv \cos\theta_i$.
The isospin part of the neutral current becomes
\begin{equation}
(\bar c_L\ \bar t_L\ \bar Q_L)
          \left[ \begin{array}{ccc}
                    C_2^2 & -S_2C_2S_3 & +S_2C_2C_3 \\
                   -S_2C_2S_3 & C_3^2 + S_2^2S_3^2 & C_2^2S_3C_3 \\
                  +S_2C_2C_3 & C_2^2S_3C_3 & S_2^2C_3^2 + S_3^2
          \end{array} \right] \gamma_\mu
          \left( \begin{array}{r}
                    c_L \\ t_L \\Q_L
          \end{array} \right).
\end{equation}
One sees that only the $Zc\bar c$ coupling is affected at tree level,
\begin{equation}
 v_c = \sqrt{\rho}~ \left[t_3^c C_2^2
  - 2 Q_c \sin^2\bar{\theta}_W\right], \ \ \ \
 a_c  = \sqrt{\rho}~ t_3^c C_2^2,
\label{coupl}
\end{equation}
where $\sqrt\rho$ and $\bar{\theta}_W$ are in standard notation.
One finds
\begin{eqnarray}
\label{Rl}
R_\ell & \simeq & R_l^{\rm SM} \left(1 - 0.41 S_2^2
                            + 0.30 S_2^4\right),  \\
\label{Rb}
R_b & \simeq & R_b^{\rm SM}/\left(1 - 0.41 S_2^2
              + 0.30 S_2^4\right), \\
\label{Rc}
R_c & \simeq & R_c^{\rm SM} \left(1 - 2.41 S_2^2 + 1.75 S_2^4\right)
        /\left(1 - 0.41 S_2^2 + 0.30 S_2^4\right).
\end{eqnarray}
Thus, interestingly, $R_c$ moves down while $R_b$ goes up!
However, $R_\ell^{\rm exp} = 20.788 \pm 0.032$
is a stringent constraint with 0.15\% accuracy.
Allowing for 2$\sigma$ variation, 
and $\alpha_S(m_Z)$ values up to 0.126,
taking $m_t = 180$ GeV and varying $m_{H^0}$
between 70 and 300 GeV,
we find that higher $\alpha_S$ and lower $m_{H^0}$ values are favored,
with $S_2^2$ varying between  0.007 and 0.009,
and $\delta R_c$, $\delta R_b$ between  $-0.0024$ and $-0.0032$,
$0.0006$ and $0.0008$, respectively.
The direction is right, but far from sufficient.
The reason can be traced to the smallness of 
$\delta R_\ell^{\rm exp.}$.
It would therefore be ideal if one could
decrease $R_c$ and increase $R_b$ substantially,
but keeping $\delta R_\ell \sim \delta R_\ell^{\rm exp.}$.
That is, some fine-tuning is needed between 
$\delta R_c$ and $\delta R_b$, and eq. (1) is not self-consistent.

In any case, this does not seem to be achieveable 
within minimal extensions of SM,
including$^{2{\scriptsize )}}$ minimal SUSY (MSSM).

\vglue 0.4cm
{\bf \noindent 3. Radical, Provocative Possibility}
\vglue 0.2cm

Maintaining one singlet quark $Q$, let us in our minds relax on the
requirement on ``minimal" extension.
One can now decouple the problem from $R_\ell$ and $\alpha_S$,
since they are not {\it intrinsically} related to $R_b$and  $R_c$.
Note that $Q$ does not affect $\ell$ and $d$ sectors directly,
and large $\delta R_c$ is clearly possible,
which just means large $S_2$.
The question now is whether $\delta R_b \sim -\delta R_c$
(so $\delta R_h\approx 0$ is maintained) is possible
with just one $Q= 2/3$ singlet quark.
It turns out that this in fact occurs {\it naturally}
in the ``light, hidden top" scenario of ref. 4,
where the top quark (doublet partner of $b$ quark) is light,
$m_{H^0} < m_t < M_W$, and the singlet quark $Q$ is the
observed heavy quark with $m_Q \simeq 180$ GeV. 
Here,
the light top hides below $M_W$ because of induced 
$t\to cH^0$ decay dominating over the
standard 3-body $t\to bW^*$ decay.$^{4{\scriptsize )}}$

The point is that this occurs only if both $S_2$ and $S_3$ are large,
as can be seen from a formula similar to eq. (4).
From eqs. (3) and (4),
both $t$ and $Q$ would now enter the $Zb\bar b$ loop vertex,
since the GIM-breaking $c$, $t$ and $Q$ mixing 
modifies the charged and neutral currents.
The leading effect can be summarized as a shift to an
{\it effective} top mass,
\begin{equation}
m_t^2 \longrightarrow (m_t^{\rm eff.})^2
                     = m_t^2 + 2S_3^2m_t(m_Q - m_t)
                      + S_3^2(S_2^2 + S_3^2 - S_2^2S_3^2)(m_Q - m_t)^2.
\end{equation}
in the SM  $Zb\bar b$ vertex.
That is, $R_b^{\rm SM}(m_t)$ in eq. (7) should be replaced by
$R_b^{\rm SM}(m_t^{\rm eff.})$.
Note that it has been known for a long time that the
``$R_b$ problem" itself can be phrased as $R_b^{\rm exp.}$
favors lighter top quark mass.
We find that if $S_2^2 + S_3^2 < 0.5$ and $S_2^2 < S_3^2$,
$m_t^{\rm eff.} < 125$ GeV.
We therefore now have a new strategy:
large $S_2$ drives down $R_c$, while
large $S_2$ and $S_3$ leads to the light top possibility
which drives down $R_b$ indirectly via loop corrections.

As an illustration of this strategy, let us take (in contrast to eq. (1))
\begin{equation}
   R_b \simeq 0.2219\;\; (+3.7\sigma\ {\rm shift}), \;\;\;
   R_c \simeq 0.1616\;\; (-1.4\sigma\ {\rm shift}).
\end{equation}
Thus, with $R_c/R_c^{\rm SM} \simeq 0.940$,
we find from eq. (8) that
\begin{equation}
S_2^2 = 0.0305,
\label{s2}
\end{equation}
which is larger than the values from previous section.
Inserting this value of $S_2$ into eq. (7), we find that
$R_b^{\rm SM}(m_t^{\rm eff.}) = 0.219$,
which implies that  $m_t^{\rm eff.} \simeq 100$ GeV.
Taking $m_t = 70$ GeV and $m_Q = 180$ GeV,
and solving eq. (9), we get
\begin{equation}
S_3^2 \simeq 0.27,
\label{s3}
\end{equation}
which is larger than $S_2^2$.
Note that $t$ is still dominantly
the $SU(2)$ partner of the $b$ quark, which justifies our flavor label.

\vglue 0.4cm
{\bf \noindent 4. CAVEATS!}
\vglue 0.2cm

Two problems emerge behind our back at this point.
First, a light top and $m_Q \simeq 180$ GeV in 
the $W$ and $Z$ two-point functions would result in too low 
a value for $\delta\rho$ (or, insufficient $\Delta T$).
In other words, $R_\ell$ comes back to haunt us in a different way.
Second,
one can check that the values of $S_2$ and $S_3$ 
from eqs. (11) and (12) are not large enough 
to support $t\to cH^0$ decay dominating over $t\to bW^*$.
It is rather amusing, however, that both problems can be
removed by the introduction of a second Higgs doublet
that does not mix very much with the standard one.
All one needs to do is to demand that
$m_{H^+} > v$ but $m_{h^0} < M_W$,
where $h^0$ stands for lightest neutral (pseudo)scalar.
In fact it is necessary to have $m_{h^0}$ as light as possible
so that $t\to ch^0$ would not have any phase space suppression.
A heavy $H^+$ does not appear strongly in loop diagrams
(such as $b\to s\gamma$ and $B^0$-$\bar B^0$
mixing),
but provides a sizable extra $\Delta T$ via $H^+ - h^0$ splitting.
A very light $h^0$ is possible if the accompanying
nonstandard neutral Higgs is heavier than $M_Z$.
Thus, we find a viable solution to $R_b$ -- $R_c$ problem
at the cost of introducing an exotic singlet charge 2/3 quark 
and a second Higgs doublet, with parameters arranged in a rather
special ``corner".

\vglue 0.4cm
{\bf\noindent 5. Phenomenological Discussion}
\vglue 0.2cm

For sake of space, let us summarize the tantalizing phenomenological
consequences of this peculiar but not unnatural solution:
\vskip -0.3cm
\begin{itemize}
\item $t\to ch^0$ is dominant over $t\to bW^*$ but not overwhelmingly so.
Thus, Tevatron should restudy the  $m_t < M_W$ region
for BR$(t\to e\nu + X)$ not much smaller than 1/18.
\item The leading $Q$ decays are
$Q\to bW$, $sW$; $tH$, $tZ$; $cH$, $cZ$, with
relative weights 66.3\%, 5.4\%, 14.6\%, 7.3\%, 3.3\%, 3.0\%,
respectively. Thus, more than 70\% of $Q$ decays contain $W$'s,
while, 
since $t\to cH^0$, $bW^*$ and $H^0 \to b\bar b$,
the $b$ quark content of $Q$ decay is close to unity.
Both are in agreement with recent CDF results.$^{5{\scriptsize)}}$
\item BR$(Z\to t\bar{c}+\bar{t}c)\sim$ few $\times 10^{-4}$. 
One should search for
the $Z\to \ell\nu bc$ signal.
\item {\it Dramatic} consequences at LEP-II:
TOPONIUM afterall!? 
Open top could also appear during the LEP 1.6 run next summer,
with an extremely light Higgs as a further bonus.

\end{itemize}

\vglue 0.4cm
{\bf\noindent 6. Conclusion: \rm
We should know within a year, 
before the $R_b$ -- $R_c$ problem itself is settled!}
\vglue 0.2cm

\vglue 0.4cm
{\bf\noindent 7. References \hfil}
\vglue 0.2cm

\end{document}